\documentclass[twoside]{ilcws07}
\usepackage[latin1]{inputenc}
\usepackage[dvips]{graphicx,epsfig,color}
\usepackage{wrapfig,rotating}
\usepackage{amssymb,amsmath,array}

\begin{document}
\title{
Model-independent WIMP Searches at the ILC} 
\author{Christoph Bartels$^{1,2}$ and Jenny List$^1$
\vspace{.3cm}\\
1- DESY - FLC \\
Notkestr. 85, 22607 Hamburg - Germany
\vspace{.1cm}\\
2- Universit\"at Hamburg - Inst. f. Experimentalphysik \\
Luruper Chaussee 149, 22761 Hamburg - Germany\\
}

\maketitle

\begin{abstract}
We investigate the possibility to detect WIMPs at the ILC in a model-independent way using events with single photons. The study is done with a full detector simulation of the LDC detector and MarlinReco. It turns out that WIMPs are observable this way at the ILC if their coupling to electrons is not too small. Beam polarisation can increase the accessible phase space significantly.
 
\end{abstract}

\section{Introduction}
Weakly interacting massive particles (WIMPs) are currently favoured as candidates for Dark Matter, which makes up about 20$\%$ percent of the total matter-energy content of the universe. Although the lightest supersymmetric particle, which is the lightest neutralino in many SUSY models, would make a very good WIMP if SUSY with R-Parity conservation is realised in nature, most extensions of the Standard Model of particle physics contain WIMPs. In fact the more general requirement is that some new conserved quantum number makes the lightest of the new particles stable. 
The International Linear Collider (ILC) offers the possibility to look for WIMPs in a rather model-independent way, which first has been pointed out in~\cite{Birkedal:2004xn}. Here, the expected sensitivity to such a WIMP signal, the achievable mass resolution and the influence of beam polarisation are studied using a full detector simulation.  

Assuming that the cosmic relic density of WIMPs is determined by pair annihilation of WIMPs, and that an unknown branching fraction $\kappa_e$ of these annihilations proceeds into an $e^+e^-$ pair($XX\rightarrow e^+e^-$), crossing relations can be used to derive an expected cross-section for the reverse process, i.e. $e^+e^-\rightarrow XX$. This cross-section contains as free parameters:

\begin{itemize}
\item the $e^+e^-$ branching fraction $\kappa_e$
\item the mass of the WIMP $M_X$
\item the spin of the WIMP $S_X$
\item the angular momentum of the annihilation's dominant partial wave $J$.
\end{itemize}

In order that this process be observable in a collider detector, where the WIMPs themselves leave no signature, an additional photon from initial state radiation is required: $e^+e^-\rightarrow XX\gamma$. 

The main Standard Model background process for this reaction is neutrino pair production, again with an ISR photon: $e^+e^-\rightarrow \nu\bar{\nu}\gamma$. At energies significantly above the $Z^0$ pole, this reaction is dominantly mediated by $t$-channel $W$-exchange and can thus be reduced significantly by choosing the appropriate polarisations for the electron and position beams. Therefore we consider three possible scenarios for the helicity structure of the WIMP coupling to electrons:

\begin{itemize}
\item the same as the SM charged current weak interaction, i.e. only $\kappa(e^-_Le^+_R)$ is nonzero
\item parity and helicity conserving, i.e. $\kappa(e^-_Le^+_R) = \kappa(e^-_Re^+_L)$
\item opposite to SM charged current weak interaction, i.e. only $\kappa_e = \kappa(e^-_Re^+_L)$ is nonzero.
\end{itemize}

Especially in the last case a significant enhancement of the signal over background ratio is expected.



\section{Software and Reconstrucion Tools}
The neutrino background has been generated with NUNUGPV~\cite{Montagna:1996ec}, which is a dedicated generator for neutrino pair production with up to three photons $e^+e^-\rightarrow \nu \bar{\nu} \gamma ( \gamma  \gamma )$. $1.2 \cdot 10^6$ events with at least one photon with an energy $E_{\gamma}$ 8~GeV $<E_{\gamma}<$ 250~GeV and a polar angle $\theta_{\gamma}$  15$^{\circ} < \theta_{\gamma} < 165^{\circ}$ were generated at a center-of-mass energy of $\sqrt{s} = 500$, corresponding to an integrated luminosity of 500~fb$^{-1}$. These events serve not only to describe the irreducible Standard Model background, but are also reweighted w.r.t. energy and polar angle of the photon according to the WIMP cross-section. The benefit of the is method is that in this way a signal sample can be obtained for all mass and spin hypotheses to be tested without applying the detector simulation again, which reduces the processing time tremendously.

This sample has been subjected to the full LDC detector simulation, using the detector model LDC01Sc with a 4~T magnetic field and Mokka~6.1~\cite{Mokka}. The events were then reconstructed with MarlinReco~\cite{Marlin}, using the WOLF~\cite{WOLF} algorithm for particle flow and a simple selection demanding  $E_{\gamma}<$ 10~GeV and 20$^{\circ} < \theta_{\gamma} < 180^{\circ}$. For the energy resolution studies described in the following, also an angular match to the generated photon was required.

\section{Energy Resolution Studies}

\begin{figure}
\centerline{
\includegraphics[width=0.5\columnwidth]{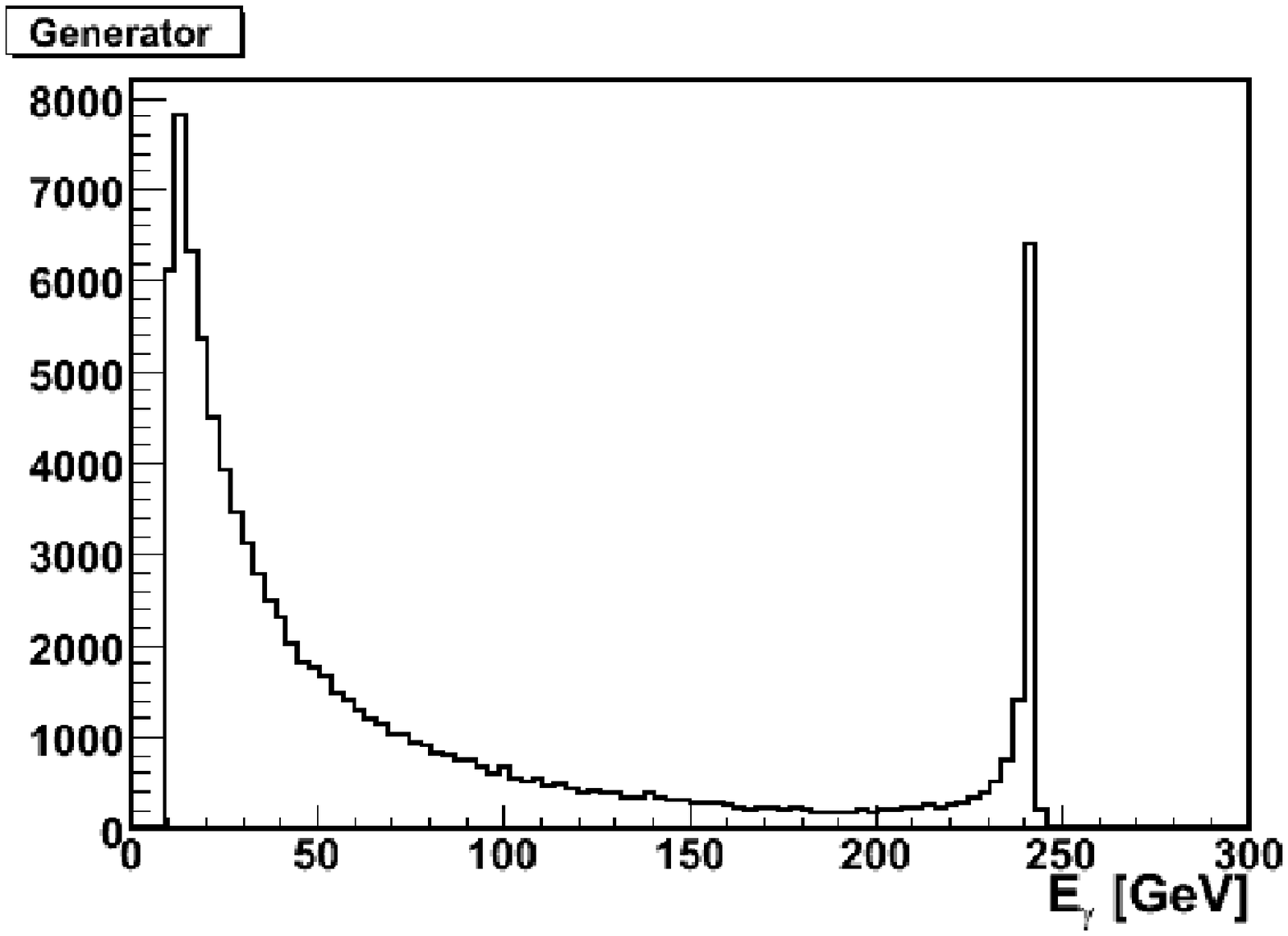}
\includegraphics[width=0.5\columnwidth]{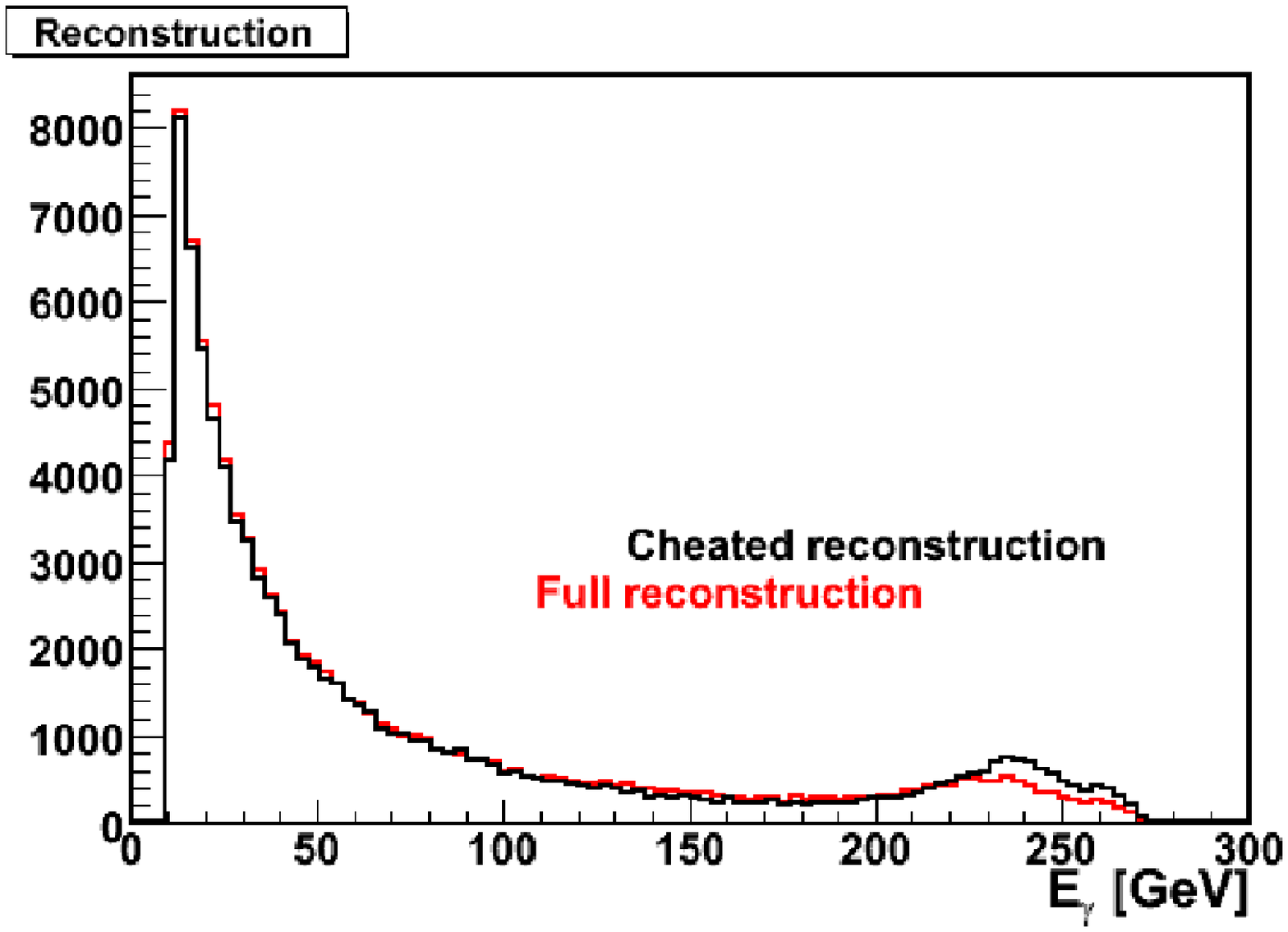}
}
\caption{Energy spectra of the event's most energetic photon at generator level (left) as well as cheated and fully reconstructed (right)}\label{Fig:Egamma}
\end{figure}

The influence of the detector's energy resolution has been studied at two levels: besides the full reconstruction, the so called cheated reconstruction makes use of the simulation information for associating calorimeter clusters with particles. This way the pure detector resolution and calibration can be disentangled from confusion effects, noise and so on. Figure~\ref{Fig:Egamma} shows the generated (left) as well as the cheated and fully reconstructed (right) energy spectra of the most energetic photon candidate of the events.

\begin{figure}[h]
\centerline{
\includegraphics[width=0.5\columnwidth]{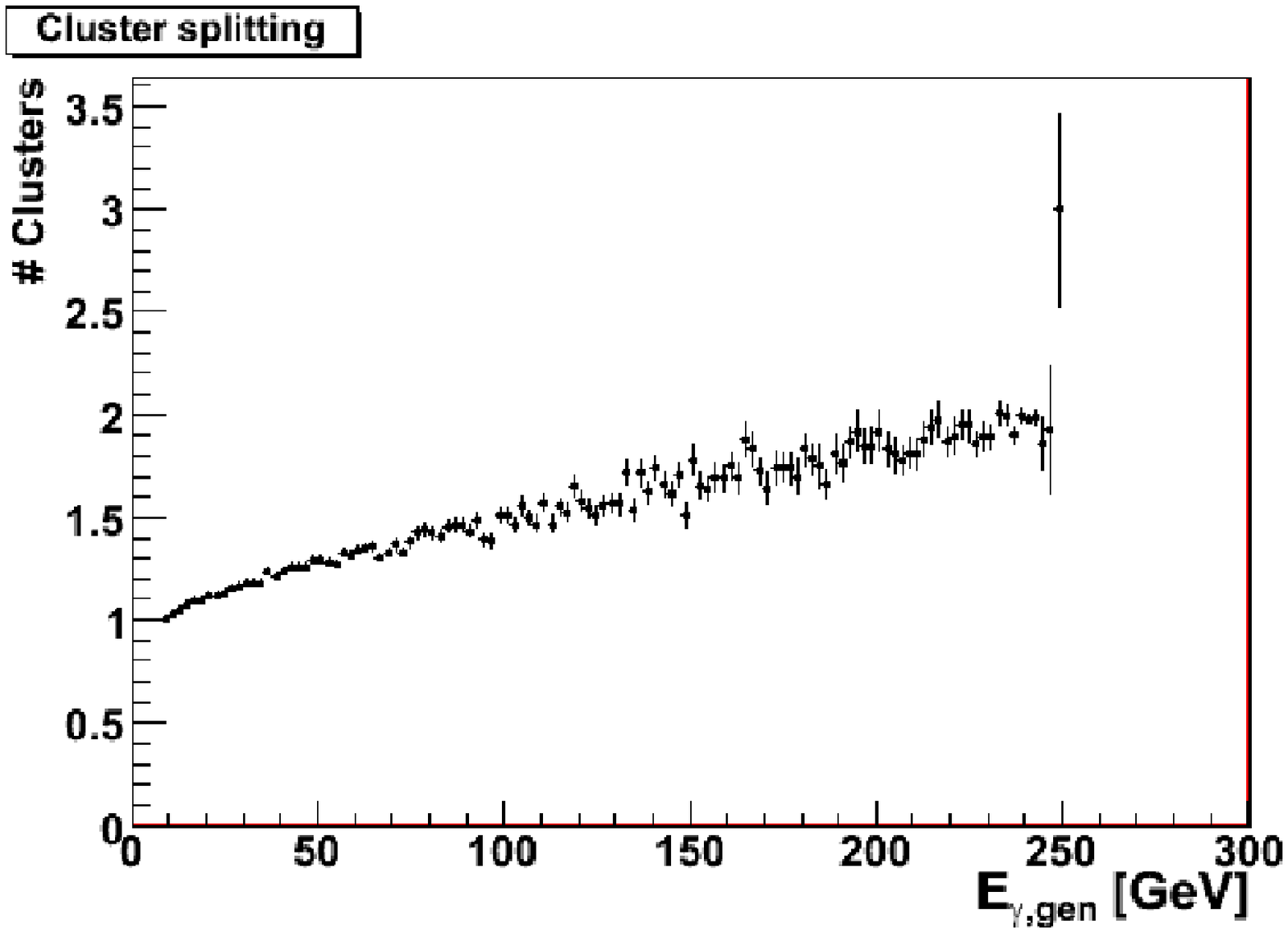}
\includegraphics[width=0.5\columnwidth]{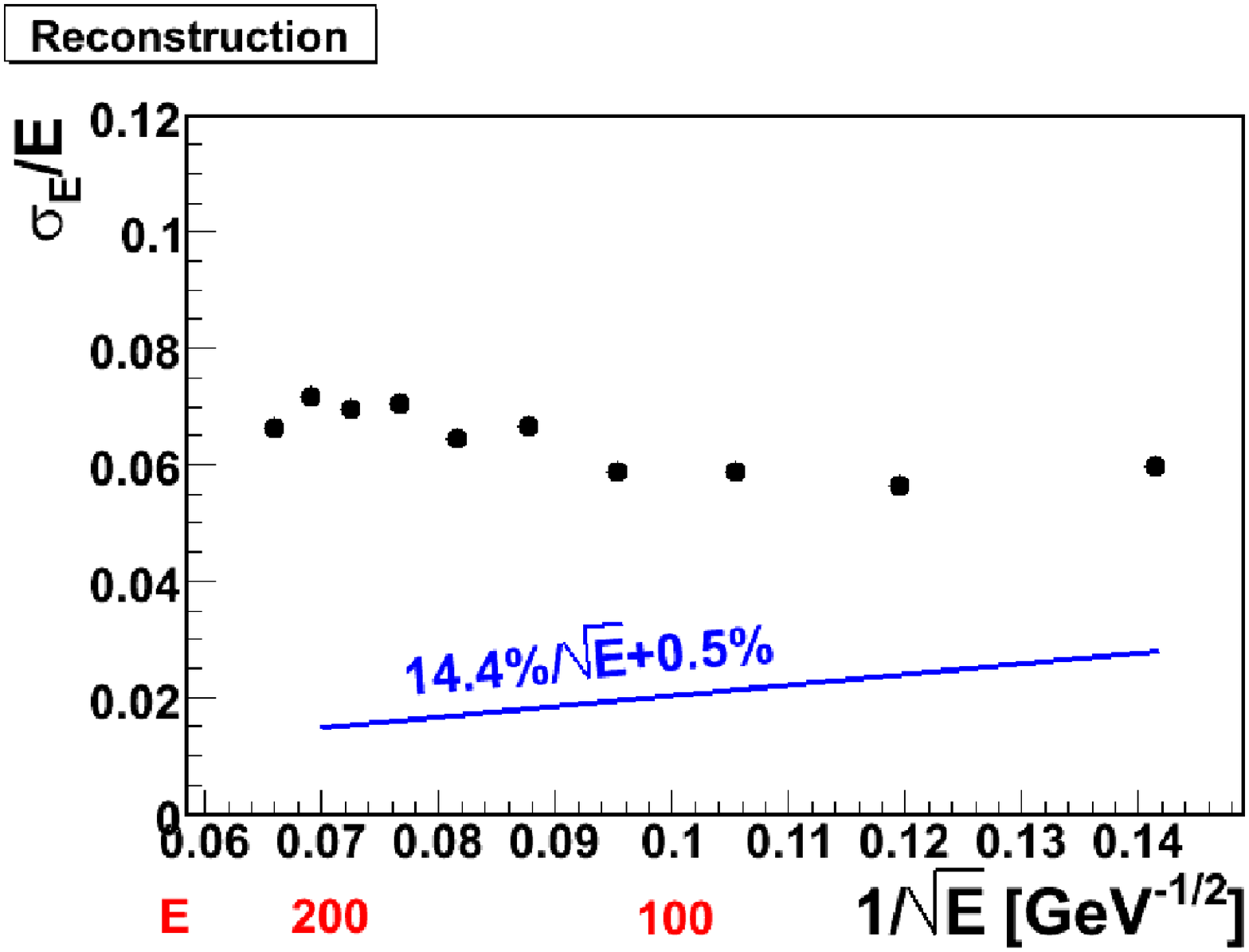}
}
\caption{Left: Mean number of reconstructed photon candidates per generated photon vs the generated photon's energy. \newline Right: Photon energy resolution after the recombination procedure vs $1/\sqrt{E}$ .}\label{Fig:NclusEres}
\end{figure}

While the peak at $E_{\gamma}=240$~GeV from radiative returns to the $Z^0$ resonance is extremely clear at the generator level, it his heavily smeared after reconstruction, even for the cheated case. The full reconstruction yields even less photons at high energies. 
The reason for this effect can be seen in the left plot of figure~\ref{Fig:NclusEres}, which shows the mean number of photon candiates per generated photon as a function of the generated photon's energy.

It is clearly visible that at high photon energies the WOLF algorithm tends to split clusters stemming from one photon into several photon candidates. Therefore we apply a merging procedure to neighboring photon candidates, after which the fully reconstructed photon energy distribution is practicaly identical to the cheated spectrum shown in the right hand plot of figure~\ref{Fig:Egamma}.

The energy resolution obtained after the recombination procedure is shown in right plot of figure~\ref{Fig:NclusEres} as a function of $1/\sqrt{E}$. With being roughly constant at about $6\%$, it seems to be significantly worse than the $14.4\%\sqrt{E}+0.5\%$ aimed for at the ILC. This effect has been meanwhile tracked down to an imperfect calibration, which means that the results presented in the next section are expected to improve further once a better calibration procedure is applied.

\section{Preliminary Analysis Results}
Currently, the following scenarios have been investigated:
\begin{itemize}
\item WIMP spin: P-wave annihilation (J=1) for $S_X = 1$ and  $S_X = \frac{1}{2}$ 
\item WIMP couplings: $\kappa(e^-_Le^+_R) > 0$, $\kappa(e^-_Re^+_L) > 0$ and $\kappa(e^-_Le^+_R) = \kappa(e^-_Re^+_L) > 0$ 
\item polarisation: unpolarised beams, $e^-$ polarisation only ($P_{e^-} = 0.8$) and additional $e^+$ polarisation ($P_{e^-} = 0.8$ and $P_{e^+} = 0.6$).
\end{itemize}

\subsection*{Observation reach}
For each combination of these parameters, the reach of the ILC with an integrated luminosity of 500~fb$^{-1}$ at $\sqrt{s}=500$~GeV for a 3$\sigma$ observation of WIMPs has been determined as a function of the WIMP mass. Due to the high irreducible background from Standard Model neutrino production, the sensitivity has been obtained statistically by using fractional event counting~\cite{Junk:1999kv} as implemented in the RooT class TLimit.

\begin{figure}[h]
\centerline{
\includegraphics[width=0.33\columnwidth]{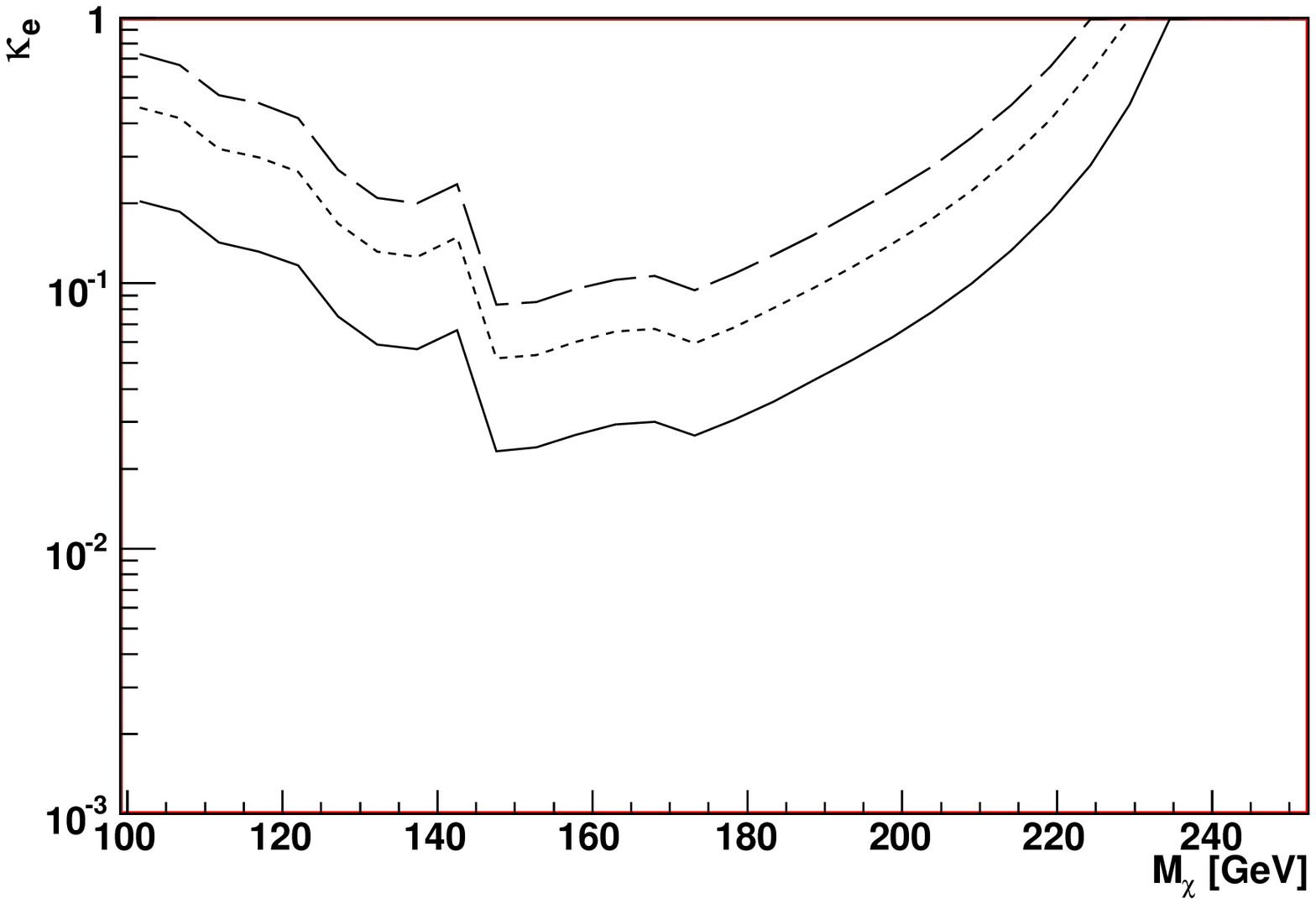}
\includegraphics[width=0.33\columnwidth]{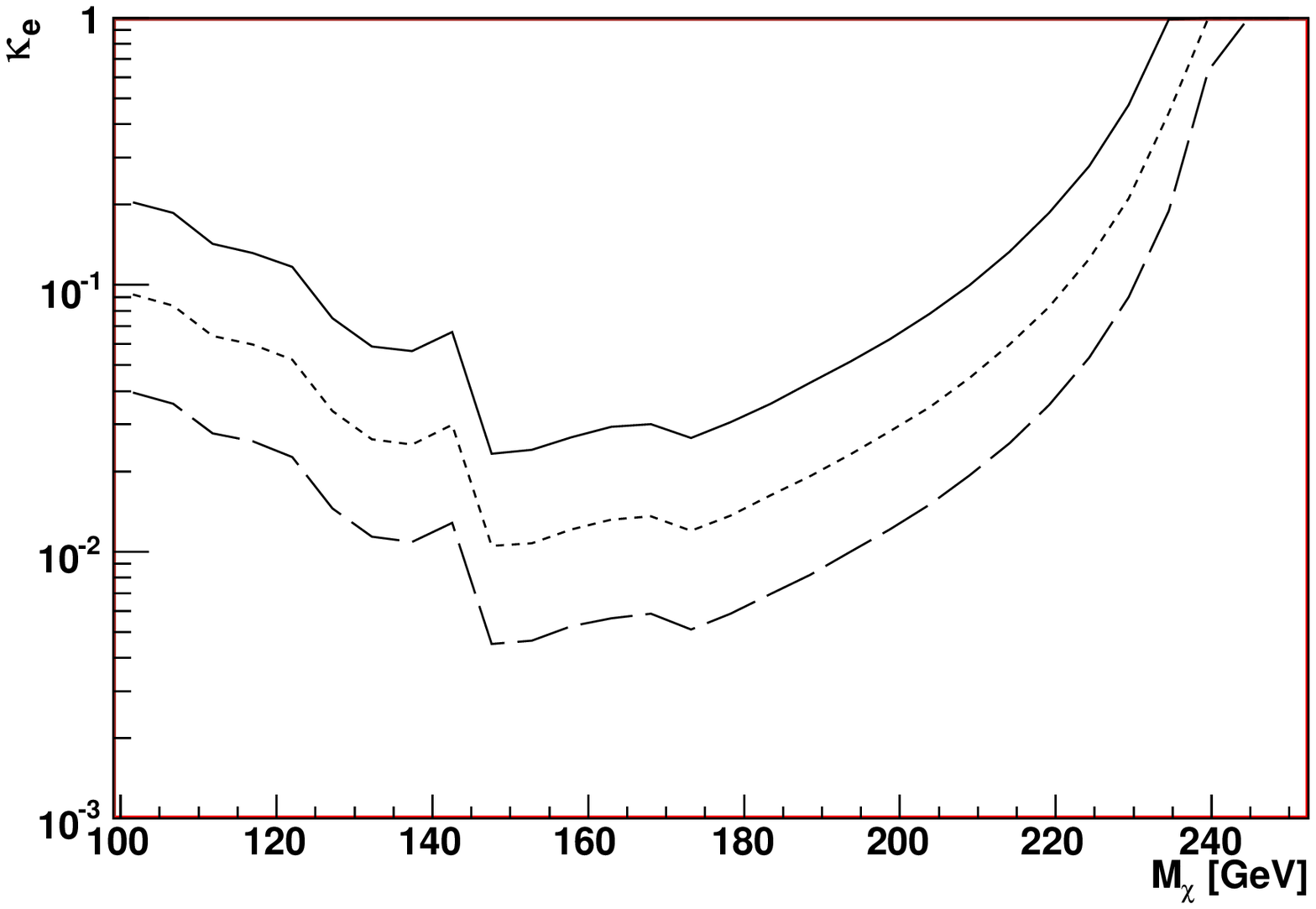}
\includegraphics[width=0.33\columnwidth]{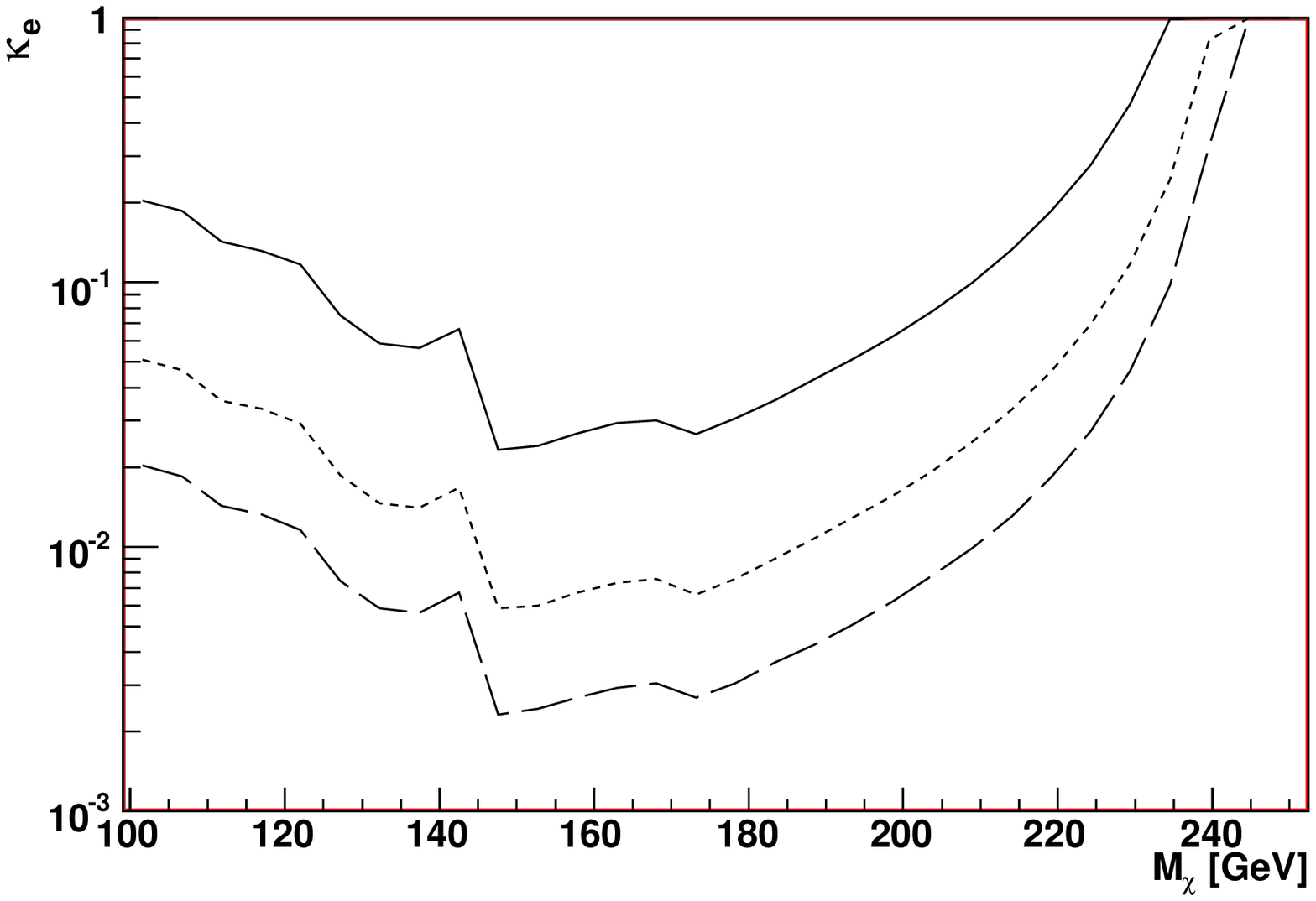}
}
\caption{$3\sigma$ observation reach of the ILC for a Spin-1 WIMP in terms of WIMP mass and $\kappa_e$ for three different assumptions on the chirality of the electron-WIMP coupling, see text. Full line: $P_{e^-} = P_{e^+} = 0$, dotted line: $P_{e^-} = 0.8, P_{e^+} = 0$, dashed line : $P_{e^-} = 0.8, P_{e^+} = 0.6$. Regions above the curves are accessible.}\label{Fig:Case1}
\end{figure}

\begin{figure}
\centerline{
\includegraphics[width=0.33\columnwidth]{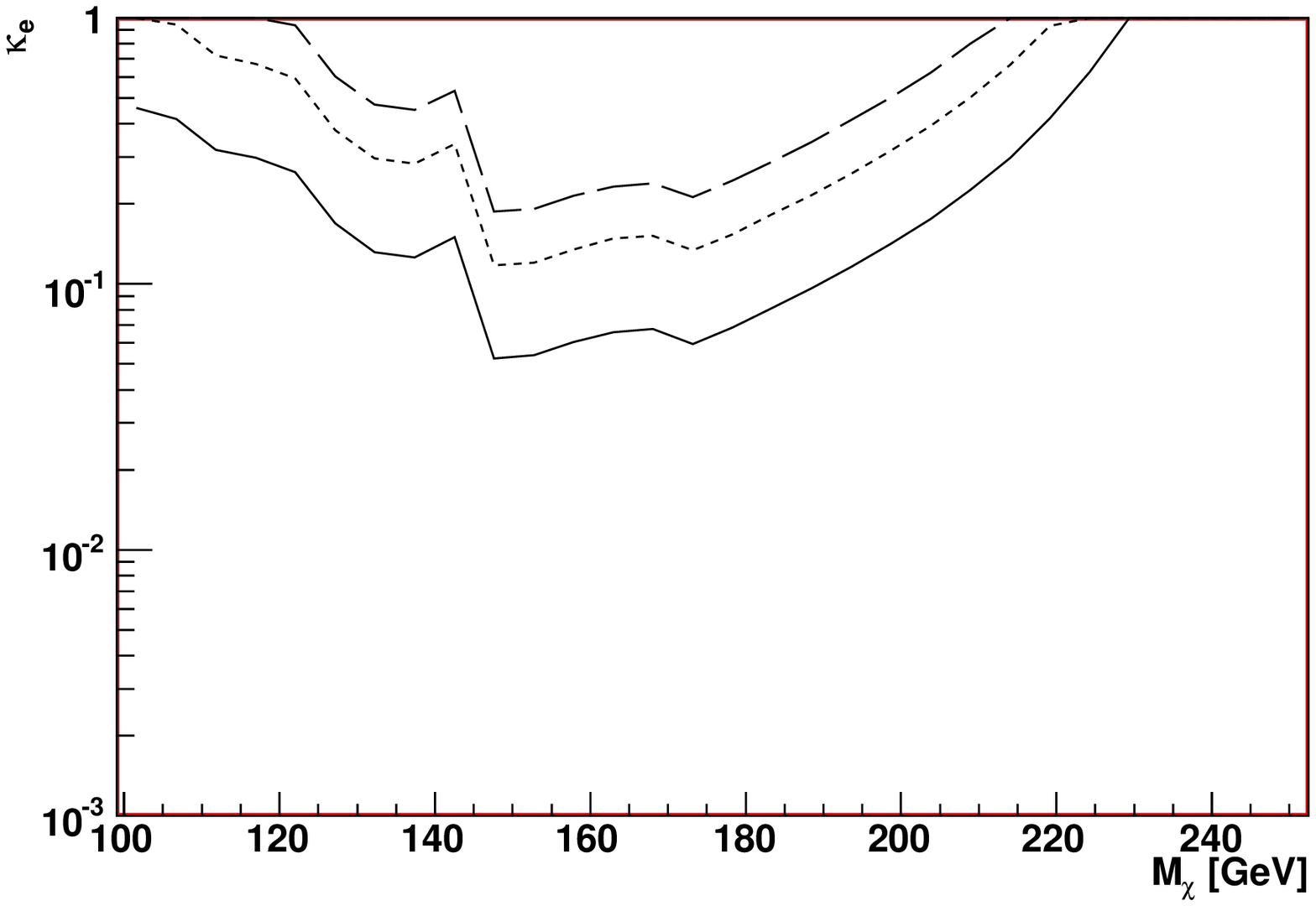}
\includegraphics[width=0.33\columnwidth]{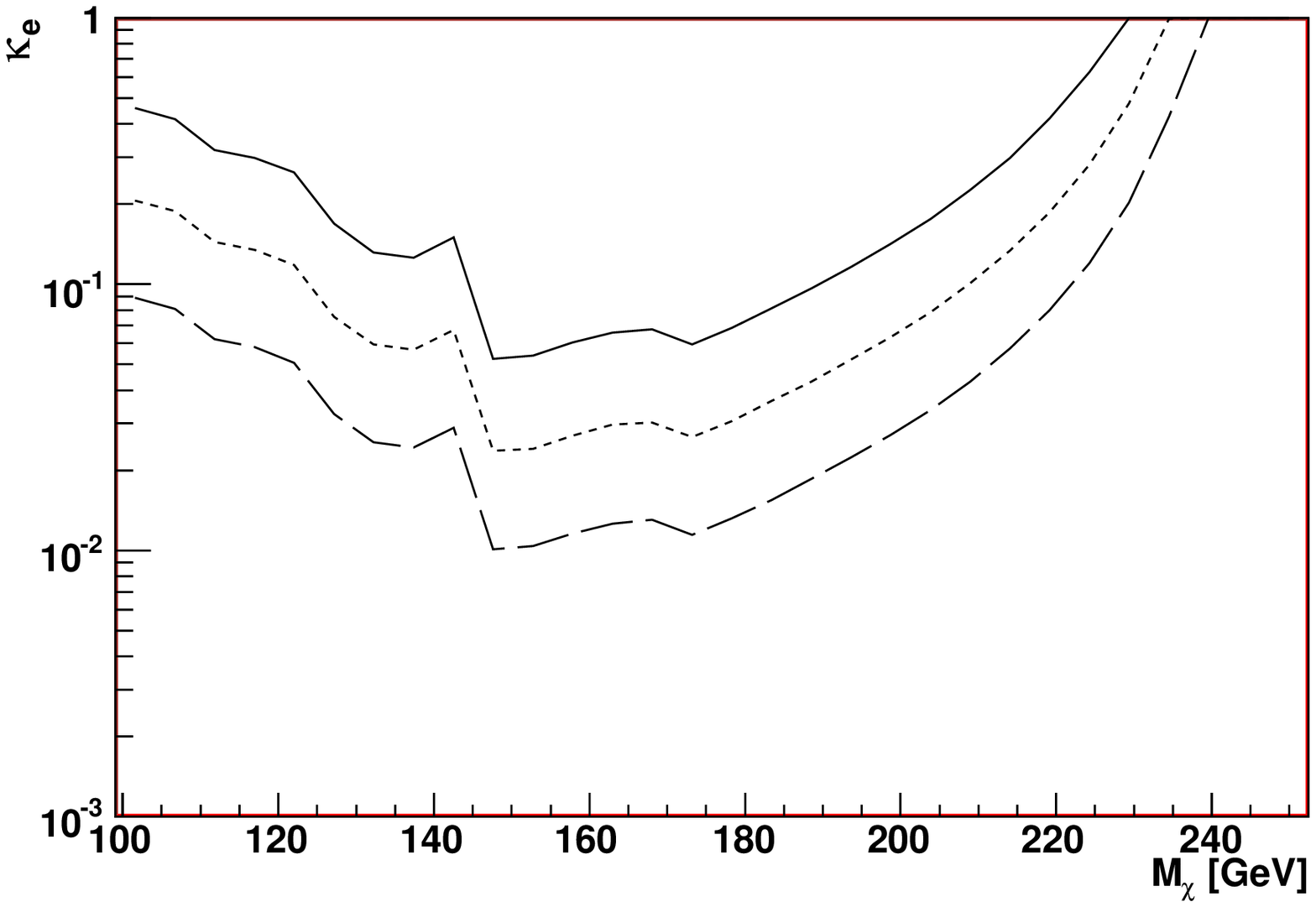}
\includegraphics[width=0.33\columnwidth]{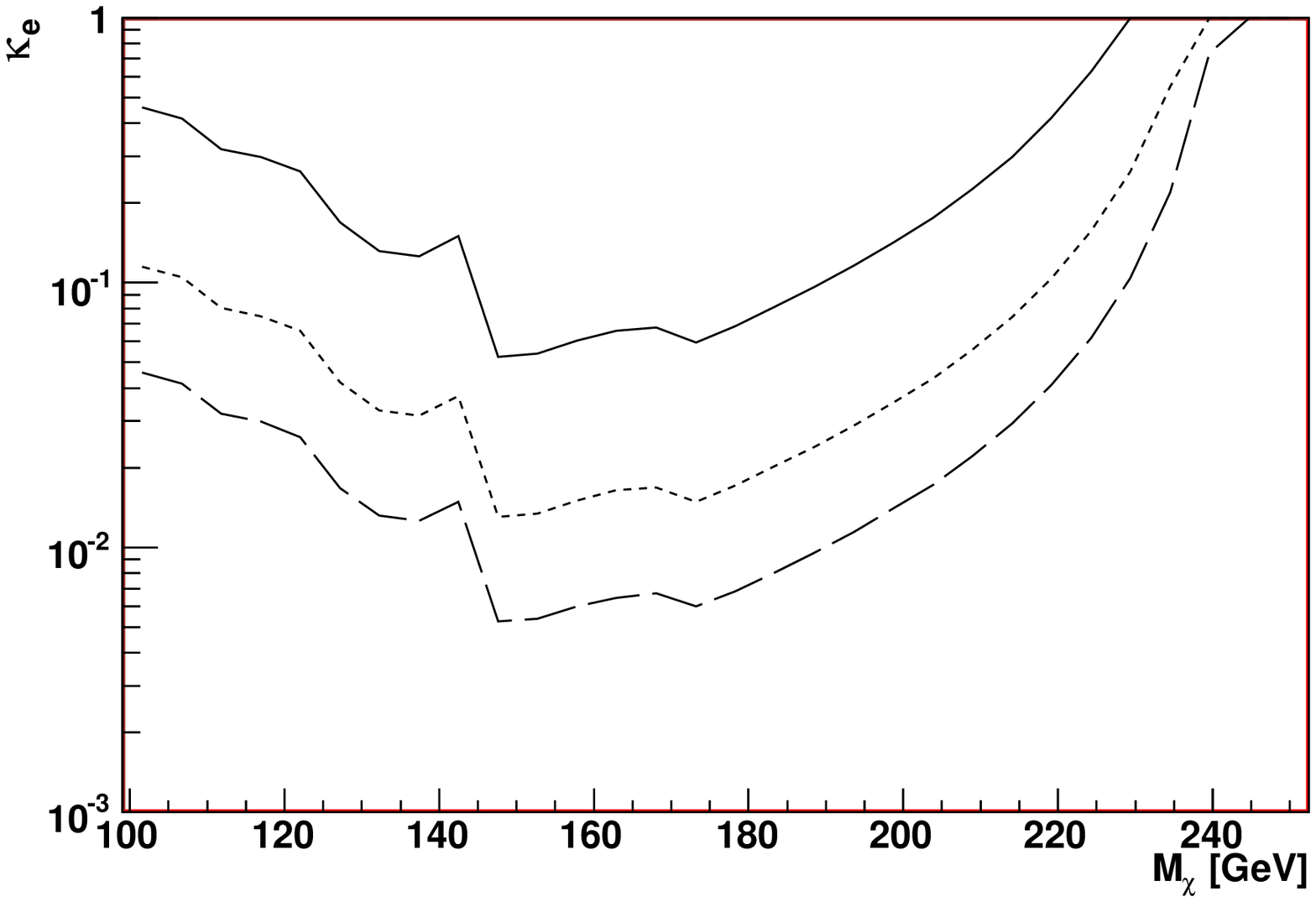}
}
\caption{$3\sigma$ observation reach of the ILC for a Spin-$\frac{1}{2}$ WIMP in terms of WIMP mass and $\kappa_e$ for three different assumptions on the chirality of the electron-WIMP coupling, see text. Full line: $P_{e^-} = P_{e^+} = 0$, dotted line: $P_{e^-} = 0.8, P_{e^+} = 0$, dashed line : $P_{e^-} = 0.8, P_{e^+} = 0.6$. Regions above the curves are accessible.}\label{Fig:Case2}
\end{figure}

Figure~\ref{Fig:Case1} shows the expected ILC sensitivity for Spin-1 WIMPs in terms of the minimal observable branching fraction to electrons $\kappa_e$ as a function of the WIMP mass. The leftmost plot shows the case where the WIMPs couple only to lefthanded electrons and righthanded positrons ($\kappa(e^-_Le^+_R)$), the middle plot shows the parity and helicity conserving case ($\kappa(e^-_Le^+_R) = \kappa(e^-_Re^+_L)$, while the right plot is dedicated to the case that the WIMPs couple to righthanded electrons and lefthanded positrons ($\kappa(e^-_Re^+_L$). The regions above the curves are accessible, where
the full line gives the result for unpolarised beams, the dotted line for $P_{e^-} = 0.8$ and the dashed line for $P_{e^-} = 0.8$ and $P_{e^+} = 0.6$. In the latter two coupling scenarios polarised beams increase the reach significantly, especially the additional positron polarisation increases the accessible range in $\kappa_e$ by about a factor of 2. 
Figure~\ref{Fig:Case2} shows the same for a Spin-$\frac{1}{2}$ WIMP. Here the sensitivity is somewhat worse, but again beam polarisation extends the observable part of the parameter space significantly.

\subsection*{Mass resolution}
If WIMPs are observed at the ILC, their mass can be determined from the recoil mass distribution of the photons:
\begin{equation}
M_{\mathrm{recoil}}^2 = s -2\sqrt{s}E_{\gamma}
\end{equation}

\begin{wrapfigure}{r}{0.5\columnwidth}
\centerline{\includegraphics[width=0.45\columnwidth]{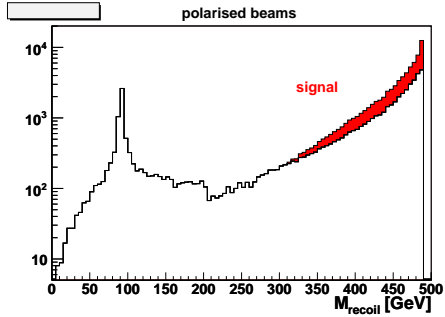}}
\caption{Recoil mass distribution for a 150~GeV Spin-1 WIMP}\label{Fig:Mass}
\end{wrapfigure}

Figure~\ref{Fig:Mass} shows an example for the recoil mass distribution for a 150~GeV Spin-1 WIMP with both beams polarised. The WIMP signal shown in dark grey kicks in at $M_{\mathrm{recoil}} =$  316~GeV.

\begin{figure}[h]
\centerline{
\includegraphics[width=0.5\columnwidth]{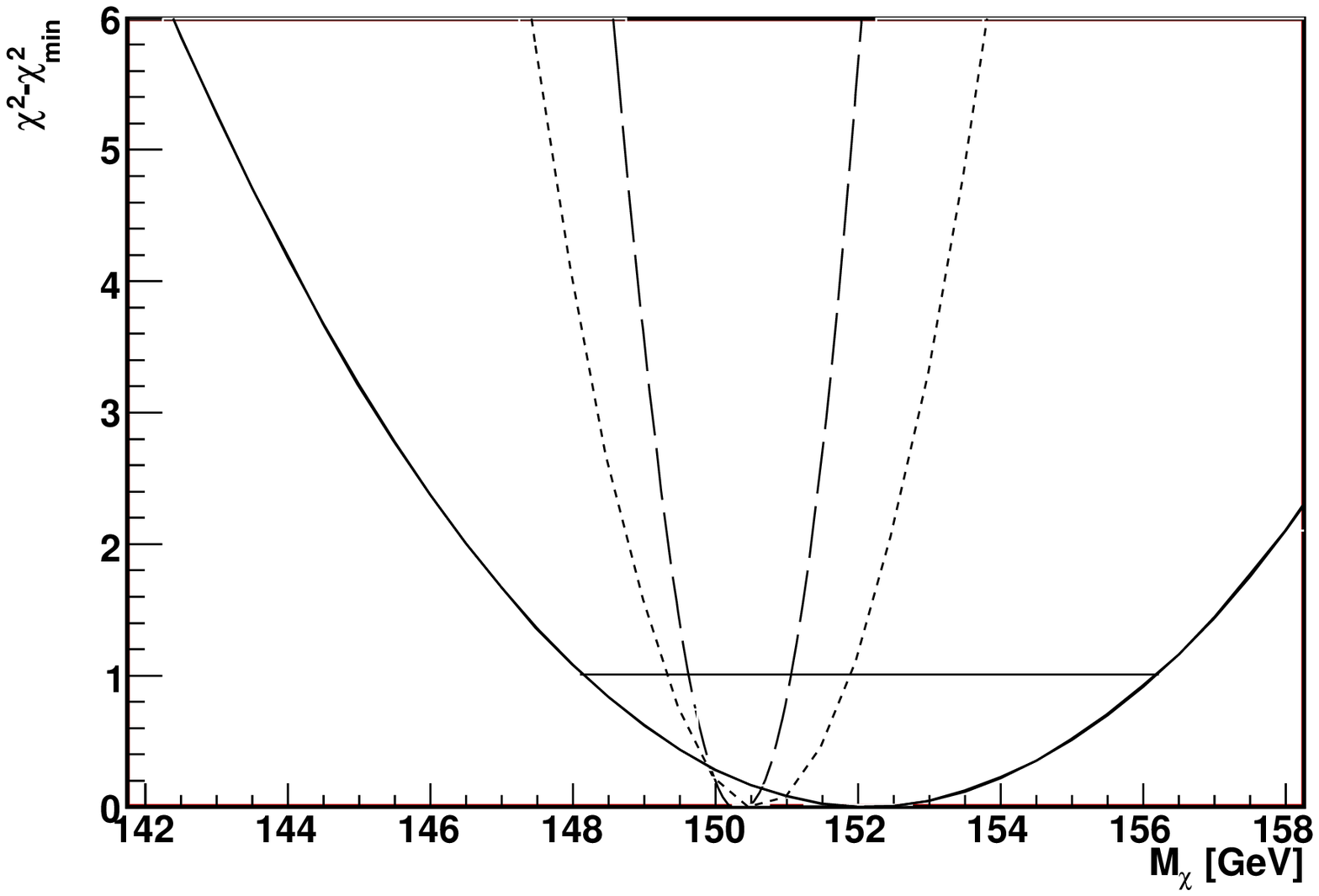}
\includegraphics[width=0.5\columnwidth]{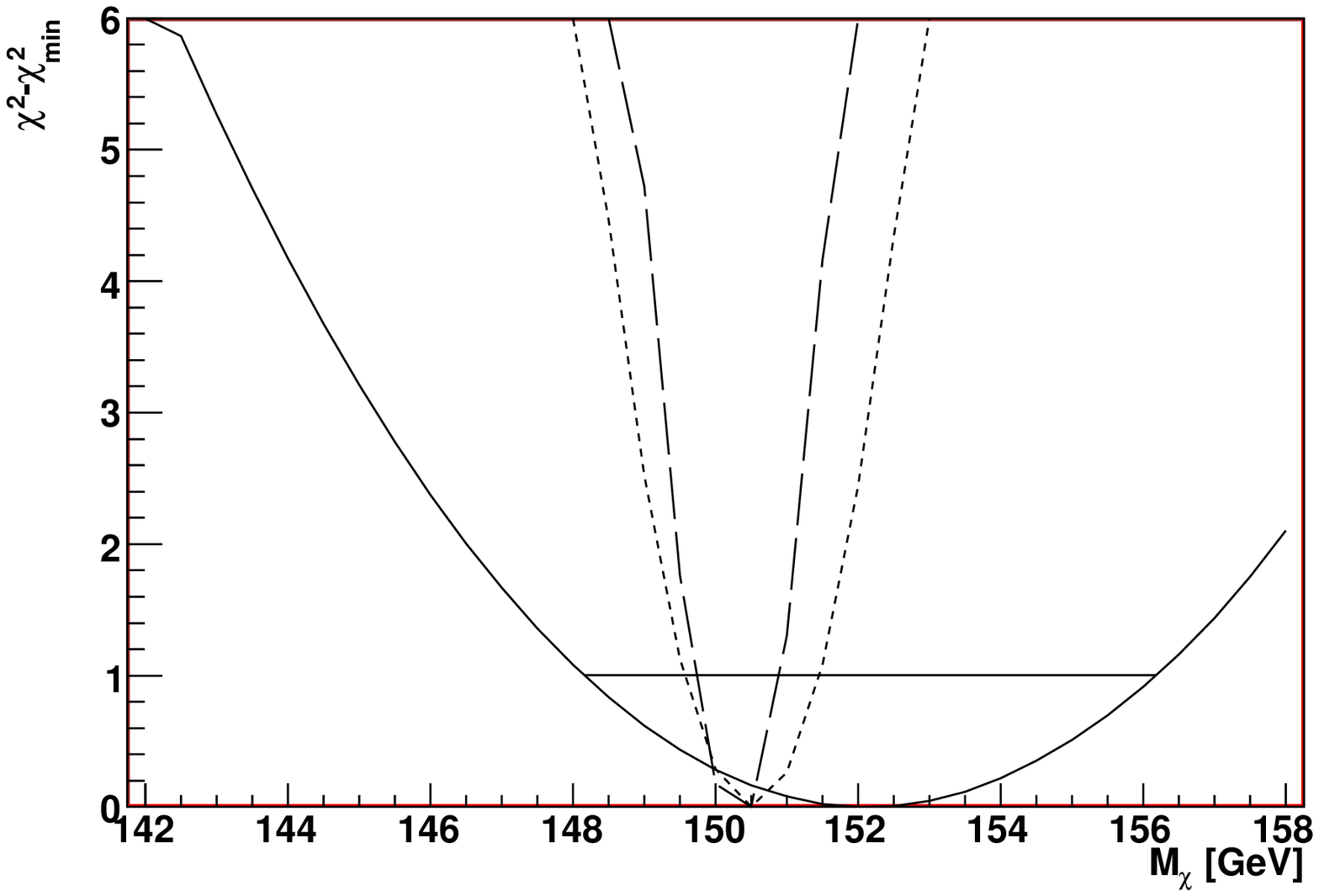}
}
\caption{$\Delta \chi^2$ from mass determination by a template method for a Spin-1 WIMP with $M_X$=150~GeV. Left: parity and helicity conserving couplings, right:$\kappa(e^-_Re^+_L)$. Full line: $P_{e^-} = P_{e^+} = 0$, dotted line: $P_{e^-} = 0.8, P_{e^+} = 0$, dashed line : $P_{e^-} = 0.8, P_{e^+} = 0.6$. }\label{Fig:Mcase1}
\end{figure}

From this distribution, the WIMP mass can be reconstructed for example by a template method. For this procedure, only 200~fb$^{-1}$ of the available MC sample have been analysed as dataset, the rest is used for the templates. Figure~\ref{Fig:Mcase1} shows the obtained $\Delta\chi^2$ as function of the reconstructed WIMP mass for a 150~GeV Spin-1 WIMP for $\kappa_e = 0.3$. The left plot shows the helicity and parity conserving case, the right plot the case  that the WIMPs couple to righthanded electrons and lefthanded positrons ($\kappa(e^-_Re^+_L$). Again the full line gives the result for unpolarised beams, the dotted line for $P_{e^-} = 0.8$ and the dashed line for $P_{e^-} = 0.8$ and $P_{e^+} = 0.6$. Without any beam polarisation, the mass resolution is about 4~GeV, which is reduced to about 1.2~GeV by switching on the electron polarisation. Positron polarisation improves the resolution by another factor 2 to about 0.6~GeV. 

\begin{figure}[h]
\centerline{
\includegraphics[width=0.5\columnwidth]{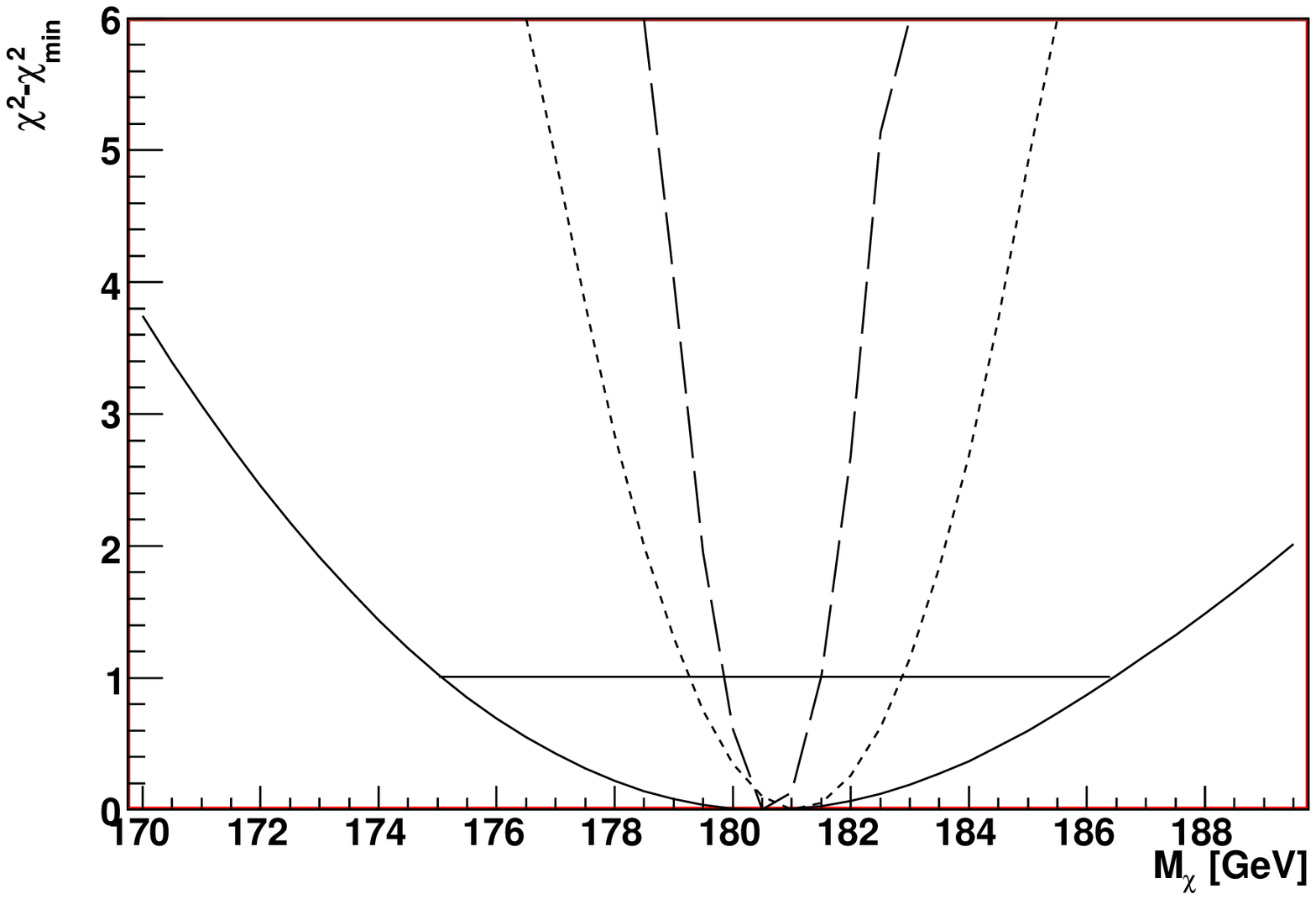}
\includegraphics[width=0.5\columnwidth]{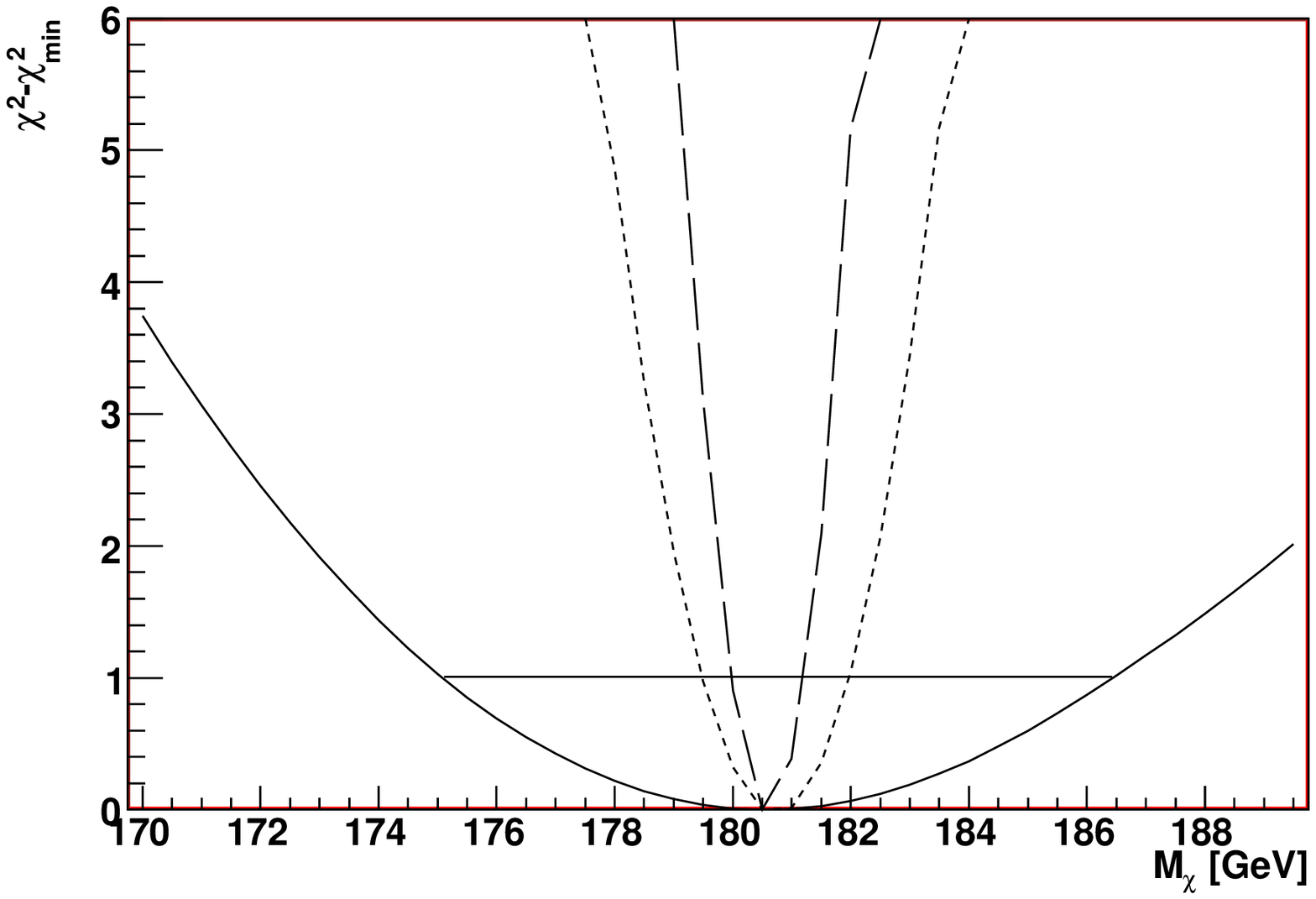}
}
\caption{$\Delta \chi^2$ from mass determination by a template method for a Spin-$\frac{1}{2}$ WIMP with $M_X$=150~GeV. Left: parity and helicity conserving couplings, right:$\kappa(e^-_Re^+_L)$. Full line: $P_{e^-} = P_{e^+} = 0$, dotted line: $P_{e^-} = 0.8, P_{e^+} = 0$, dashed line : $P_{e^-} = 0.8, P_{e^+} = 0.6$.}\label{Fig:Mcase2}
\end{figure}

Figure~\ref{Fig:Mcase2} shows the analoguos results for a 180~GeV Spin-$\frac{1}{2}$ WIMP.
As for the observation reach, the situation is slightly worse than in the Spin-1 case, but again the use of beam polarisation leads to a significant gain in resolution.

\section{Conclusions and Outlook}
The study of model-independent WIMP production at the ILC presented here is one of the first examples of analyses performed using the full simulation of the LDC detector. A reconstruction for these fully simulated events exists and usable for analysis, but needs further improvements in parallel to the optimization of the detector concept.
Already with the current level of sophistication this study shows that there is a good chance to detect WIMPs in this model-independent way at the ILC and to measure their mass with a precision of about 1~GeV. Both the range in phase space as well as the mass resolution improve significantly when polarised beams are assumed. Typically the use of 80\% electron polarisation gives improvements of a factor of two over unpolarised beams, whereas an additional positron polarisation of 60\% yields another factor of two.

The results presented here will be improved in the near future by applying a more appropriate detector calibration and by using other particle flow algorithms and photon finders. Furthermore reducible backgrounds as well as beamstrahlung have to be included in the study, before finally different detector concepts can be compared.

\section*{Acknowledgments}
The authors acknowledge the support by DFG grant Li 1560/1-1.


\begin{footnotesize}


\end{footnotesize}


\end{document}